\documentclass[english]{revtex4}
\usepackage[T1]{fontenc}
\usepackage[latin1]{inputenc}

\makeatletter

\usepackage{babel}
\makeatother
\begin{document}

\title{On the First Law of Thermodynamics for (2+1) Dimensional Charged
BTZ Black Hole and Charged de Sitter Space}

\author{Eduard Alexis Larrañaga Rubio}

\email{eduardalexis@gmail.com}

\affiliation{National University of Colombia}

\affiliation{National Astronomical Observatory (OAN)}

\begin{abstract}
In this paper we will show that using the cosmological constant as
a new thermodynamical state variable, the differential and integral
mass formulas of the first law of thermodynamics for asymptotic flat
spacetimes can be extended to (2+1) dimensional charged BTZ black
holes and charged de Sitter space.
\end{abstract}
\maketitle

\section{Introduction}

Bekenstein and Hawking showed that black holes have non-zero entropy
and that they emit a thermal radiation that is proportional to its
surface gravity at the horizon. These two quantities are related with
the mass through the identity

\begin{equation}
dM=TdS,\end{equation}

that is called \emph{the first law of balck hole thermodynamics}\cite{hawking,bekenstein}.But
when the black hole has other properties as angular momentum $\mathbf{J}$
and electric charge $Q$, the first law is generalized to 

\begin{equation}
dM=TdS+\Omega dJ+\Phi dQ,\end{equation}

where $\Omega=\frac{\partial M}{\partial J}$ is the angular velocity
and $\Phi=\frac{\partial M}{\partial Q}$ is the electric potential.
The corresponding integral Bekenstein-Smarr mass formula, and is given
by

\begin{equation}
M=\Phi Q+\frac{\left(D-2\right)}{\left(D-3\right)}\Omega J+\frac{\left(D-2\right)}{\left(D-3\right)}TS.\label{eq:bekensteinsmarr}\end{equation}

Gauntlett et. al.\cite{gauntlett} have proved that both, the differential
and integral expressions, hold for asymptotically flat spacetimes
with any dimension $D\geq4$ . For rotating black holes in anti-deSitter
spaces, Gibbons et. al.\cite{gibbons} have shown that the differential
expression hold for $D\geq4$, but the integral expresion is not satisfied.
To rectify this situation Caldarelli et. al. \cite{caldarelli} use
a cosmological constant considered as a new thermodynamical variable,
and recently, Wang et. al. \cite{wang} use this idea to show that
differential and integral expressions are valid also in $\left(2+1\right)$
dimensions for BTZ black holes with angular momentum and Kerr-de Sitter
spacetimes.

In this paper we will consider the charged BTZ black hole in $\left(2+1\right)$
dimensions to show that considering a cosmological constant as a thermodynamical
state variable, both the differential and integral mass formulas of
the first law cna be extended.

\section{The first law for the charged BTZ Black Hole}

The charged BTZ black hole \cite{martinez} is a solution of $\left(2+1\right)$
dimensional gravity with a negative cosmological constant $\Lambda=-\frac{1}{l^{2}}$.
Its line element can be written as

\begin{equation}
ds^{2}=-\Delta dt^{2}+\frac{dr^{2}}{\Delta}+r^{2}d\varphi^{2},\end{equation}

where the lapse function is

\begin{equation}
\Delta=-M+\frac{r^{2}}{l^{2}}-\frac{Q^{2}}{2}\ln\left(\frac{r}{l}\right).\end{equation}

This solution has two horizons given by the condition $\Delta=0$.
The mass of the black hole can be written in terms of the event horizon
$r_{H}$ as\cite{larr}

\begin{equation}
M=\frac{r_{H}^{2}}{l^{2}}-\frac{Q^{2}}{2}\ln\left(\frac{r_{H}}{l}\right).\end{equation}

The Bekenstein-Hawking entropy associated with the black hole is twice
the perimeter of the horizon,

\begin{equation}
S=4\pi r_{H},\end{equation}
and therefore, the mass can be written as

\begin{equation}
M=\frac{S^{2}}{16\pi^{2}l^{2}}-\frac{Q^{2}}{2}\ln\left(\frac{S}{4\pi l}\right).\label{eq:massformula}\end{equation}

If we consider an invariable $l$, this expression let us calculate
the surface gravity, temperature and electric potential for the black
hole as

\begin{eqnarray}
\kappa & = & \frac{1}{2}\left.\frac{\partial\Delta}{\partial r}\right|_{r=r_{H}}=\frac{1}{2}\left[\frac{2r_{H}}{l^{2}}-\frac{Q^{2}}{2r_{H}}\right]\label{eq:gravity}\\
T & = & \frac{\kappa}{2\pi}=\left.\frac{\partial M}{\partial S}\right|_{Q,l}=\frac{2S}{16\pi^{2}l^{2}}-\frac{Q^{2}}{2S}\label{eq:temperature}\\
\Phi & = & \left.\frac{\partial M}{\partial Q}\right|_{S,l}=-Q\ln\left(\frac{S}{4\pi l}\right).\label{eq:potential}\end{eqnarray}

Varying the mass formula (\ref{eq:massformula}) we obtain the first
law of black hole thermodynamics in differential form, 

\begin{equation}
dM=TdS+\Phi dQ.\end{equation}
But if we use equations (\ref{eq:temperature}) and (\ref{eq:potential}),
the mass formula can be written as

\begin{equation}
M=\frac{1}{2}TS+\frac{1}{2}\Phi Q+\frac{1}{4}Q^{2},\end{equation}
that does not correspond to the Bekenstein-Smarr formula (\ref{eq:bekensteinsmarr})
with $D=3$. Note that the product $TS$ in this case does not vanish, 

\begin{equation}
TS=\frac{2S^{2}}{16\pi^{2}l^{2}}-\frac{Q^{2}}{2}=\frac{2r_{H}^{2}}{l^{2}}-\frac{Q^{2}}{2},\end{equation}
except in the extremal case $Q=\frac{2r_{H}}{l}$. Otherwise, we can
try to rectify this situation by considering the effect of a varying
cosmological constant.

When considering the cosmological constant as a new state variable,
the first law in differential and integral forms must be corrected
to be

\begin{eqnarray*}
dM & = & TdS+\Phi dQ+\Theta dl\\
0 & = & TS+\Theta l,\end{eqnarray*}

where $\Theta$ is the generalized force conjugate to the parameter
$l$. This generalized force is given by

\begin{eqnarray}
\Theta & = & \left.\frac{\partial M}{\partial l}\right|_{S,Q}=-\frac{2S^{2}}{16\pi^{2}l^{3}}+\frac{Q^{2}}{2l}\\
\Theta & = & -\frac{2r_{H}^{2}}{l^{3}}+\frac{Q^{2}}{2l}.\end{eqnarray}

All the thermodynamical quantities, including this last expression
coincide with the reported results\cite{larr,wang}.

\section{The first law for the charged de Sitter space}

Now we will turn our attention to the charged de Sitter space for
wich we have a positive cosmological constant $\Lambda=\frac{1}{l^{2}}$,
and a line element 

\begin{equation}
ds^{2}=-\Delta dt^{2}+\frac{dr^{2}}{\Delta}+r^{2}d\varphi^{2},\end{equation}

where the lapse function is now

\begin{equation}
\Delta=M-\frac{r^{2}}{l^{2}}-\frac{Q^{2}}{2}\ln\left(\frac{r}{l}\right).\end{equation}

This time the ADM mass can be written in terms of the event horizon
$r_{H}$ as

\begin{equation}
M=\frac{r_{H}^{2}}{l^{2}}+\frac{Q^{2}}{2}\ln\left(\frac{r_{H}}{l}\right).\end{equation}

The Bekenstein-Hawking entropy associated with this event horizon
is again 

\begin{equation}
S=4\pi r_{H},\end{equation}
and therefore, the mass can be written now as

\begin{equation}
M=\frac{S^{2}}{16\pi^{2}l^{2}}+\frac{Q^{2}}{2}\ln\left(\frac{S}{4\pi l}\right).\label{eq:desittermass}\end{equation}

Taking first an invariable $l$, this expression gives the surface
gravity, temperature and electric potential,

\begin{eqnarray}
\kappa & = & \frac{1}{2}\left.\frac{\partial\Delta}{\partial r}\right|_{r=r_{H}}=\frac{1}{2}\left[\frac{2r_{H}}{l^{2}}+\frac{Q^{2}}{2r_{H}}\right]\label{eq:desittergravity}\\
T & = & \frac{\kappa}{2\pi}=\left.\frac{\partial M}{\partial S}\right|_{Q,l}=\frac{2S}{16\pi^{2}l^{2}}+\frac{Q^{2}}{2S}\label{eq:desittertemperature}\\
\Phi & = & \left.\frac{\partial M}{\partial Q}\right|_{S,l}=Q\ln\left(\frac{S}{4\pi l}\right).\label{eq:desitterpotential}\end{eqnarray}

Varying the mass formula (\ref{eq:desittermass}) we obtain the first
law in differential form, 

\begin{equation}
dM=TdS+\Phi dQ.\end{equation}
And using the equations (\ref{eq:desittertemperature}) and (\ref{eq:desitterpotential}),
we can write the mass formula as

\begin{equation}
M=\frac{1}{2}TS+\frac{1}{2}\Phi Q-\frac{1}{4}Q^{2},\end{equation}
that again does not correspond to the Bekenstein-Smarr general formula
(\ref{eq:bekensteinsmarr}) with $D=3$. This time the product $TS$
is 

\begin{equation}
TS=\frac{2S^{2}}{16\pi^{2}l^{2}}+\frac{Q^{2}}{2}=\frac{2r_{H}^{2}}{l^{2}}+\frac{Q^{2}}{2},\end{equation}
and it is important to note that in this case it never vanishes. However,
in order to generalize the Bekenstein-Smarr formula to this case we
will consider again the effect of a varying cosmological constant.

Now, the first law in differential and integral forms must be corrected
to be

\begin{eqnarray*}
dM & = & TdS+\Phi dQ+\Theta dl\\
0 & = & TS+\Theta l,\end{eqnarray*}

where the generalized force $\Theta$ is given by

\begin{eqnarray}
\Theta & = & \left.\frac{\partial M}{\partial l}\right|_{S,Q}=-\frac{2S^{2}}{16\pi^{2}l^{3}}-\frac{Q^{2}}{2l}\\
\Theta & = & -\frac{2r_{H}^{2}}{l^{3}}-\frac{Q^{2}}{2l}.\end{eqnarray}

\section{Conclusion}

By considering the cosmological constant as a new thermodynamical
state variable, we have generalized the integral Bekenstein-Smarr
mass formula to the cases of $\left(2+1\right)$ dimensional charged
BTZ black holes and charged de Sitter space, while the differential
form of the first law is also applicable in these cases. We also obtained
the generalized force associated with the cosmological term and the
correspondient mass formulas for these cases. It should be noted that
this results can be applicable also in higher dimensional cases.


\begin{thebibliography}{2}
\bibitem{wang}Wang, S., Wu, S-Q., Xie, F. and Dan, L. Chyn. Phys.
Lett. 23, 1096 (2006) \textbf{hep-th/0601147}

\bibitem{larr}Larranaga E.A. \textbf{0706.1599 {[}gr-qc]}

\bibitem{hawking}Hawking, S. W. Commun. Math. Phys. \textbf{43},
199 (1975)

\bibitem{bekenstein}Bekenstein, J. D. Phys. Rev. D\textbf{7}, 2333
(1973); Phys. Rev. D\textbf{9}, 3293 (1974)

\bibitem{gauntlett}Gauntlett, J., Myers, R. and Townsend, P. Class.
Quantum. Grav. \textbf{16}, 1. (1999)

\bibitem{gibbons}Gibbons, G., Perry, M.J., and Pope, C.N. Class.
Quantum. Grav. \textbf{22}, 1503. (2005)

\bibitem{caldarelli}Caldarelli, M., Cognola, G. and Klemm, D. Class.
Quantum. Grav. \textbf{17}, 399 (2000)

\bibitem{martinez}Martinez, C., Teitelboim, C. and Zanelli, J. Phys.
Rev. \textbf{D61}, 104013 (2000)
\end{thebibliography}
\end{document}